\documentclass[prb,twocolumn,amsmath,amssymb,floatfix,showpacs,superscriptaddress]{revtex4}

\usepackage{graphicx}
\usepackage{times}
\usepackage{bm}
\begin{document}

\title{Effects of coupling to vibrational modes on the ac conductance of molecular junctions}

\author{A. Ueda}
 \email{akiko@bgu.ac.il}

 \affiliation{Department of Physics, Ben Gurion University,
Beer Sheva 84105, Israel}
 \author{O. Entin-Wohlman}

\altaffiliation{Also at Tel Aviv University, Tel Aviv 69978,
Israel}

\affiliation{Department of Physics, Ben Gurion University, Beer
Sheva 84105, Israel}

\author{A. Aharony}

\altaffiliation{Also at Tel Aviv University, Tel Aviv 69978,
Israel}
\affiliation{Department of Physics, Ben Gurion University, Beer
Sheva 84105, Israel}

\begin{abstract}

We theoretically examine the effect of the coupling of the transport
electrons to a vibrational mode of the molecule on the  ac linear-response conductance
of molecular junctions. Representing the molecule by a single electronic state,
we find that at very low temperatures the  frequency-dependent
conductance is mainly enhanced (suppressed) by the
electron-vibration interaction when the chemical potential is
below (above) the energy of that  state.
The vertex corrections of the electron-vibration interaction
induce an additional peak structure in the conductance, which can
be observed by tuning the tunnel couplings with the leads.
\end{abstract}

\pacs{71.38.-k, 73.63.Kv, 73.21.La}

\maketitle

\section{Introduction}

\label{INTRO}

The fabrication of junctions made of small  molecules
 seems by now well established. \cite{reed, reichert, zhitenev, kubatkin, kushmerick,
 qui, park, leroy, smit, sapmaz, tal, leturcq}
 The transport properties of such molecular junctions are largely determined by
 the interplay between the electrical and
 the vibrational degrees of freedom. These include the couplings of the bridge with the leads,
 as well as the resonance energies. Their effects on the dc conductance of such
 junctions have been examined in many works (see below).
 Here we study the ac conductance,
 concentrating on the modulation of the frequency dependence
 brought about by the coupling to the vibrational modes. Our calculation
 is confined to the linear-response regime, and is carried out to second
 order in the coupling with the vibrational modes.

The simplest model for describing this system assigns a single
electronic level,  $\varepsilon^{}_0$,  to the molecule,  which is
connected via two leads to two electronic reservoirs. Those are
kept at different chemical potentials. When electrons pass through
the molecule, they are coupled to its vibrational modes. Since the
molecule has a finite size, those modes have finite frequencies;
in the simplest approach and at very low temperatures,  they may
be represented by a single vibrational frequency, $\omega^{}_0$,
treated within the Einstein model. Various approximate numerical
and analytical schemes have been developed for treating this
model. The dc current-voltage characteristics have been studied
 by the perturbation theory, \cite{mitra, ryndyk, egger, riwar, ora, hod, galperin}
 the rate equations,\cite{koch} and Monte Carlo methods.\cite{muhlbacher}
 Furthermore, the shot noise and the full counting statistics of the charge passing the junction
 have been studied by perturbation theory,
\cite{galperin2, haupt, avriller,schmidt, urban} rate equations,
\cite{avriller2} and the polaron approximation.\cite{zhu, maier}

In the linear-response regime, the Breit-Wigner resonance of the
dc conductance (as a function of the equilibrium chemical potential $\mu$,
and at very low temperatures) is {\em narrowed down}
by the electron-vibration (e-v) interaction, due
to the renormalization of the tunnel coupling between the molecule
and the leads (the Frank-Condon blockade). \cite{koch, ora}
However, the e-v interaction {\em does not} yield side peaks in
the conductance at energies corresponding to the molecule vibrational
frequencies, as long as it is not large enough to induce
polaronic phase transitions. For such peaks to arise
(at zero temperature and at small values of the e-v coupling), the electron needs to lose
an energy $\omega^{}_0$ (we use $\hbar=1$)  in order to excite a
vibrational mode, and this is not possible in the linear-response
regime.  Such side peaks will appear (at zero temperature) in the  {\em nonlinear-}response regime,
where the finite voltage, $V$,  allows for the opening of  inelastic channels when $eV$
exceeds $\omega^{}_{0}$ .\cite{mitra,ora} Side peaks  can be induced
by a finite temperature, but their height will be minute,
reflecting the number of available vibrational modes. \cite{ora}

The ac conductance of molecular junctions, in particular the effects of the e-v
interaction on its frequency dependence, has been studied to lesser extent.
Recently, Kubala and Marquardt \cite{kubala} wrote down
expressions for the ac conductance of  interacting electrons
traversing molecular junctions,  in the
case where the dc voltage is finite,  while the ac one is very small,
including as an example the e-v interaction.
Our expressions (which are derived by a
different approach) agree with theirs in the zero-dc bias limit. In addition,
we give details of an extra diagram
[(d) in Fig. \ref{fig:fig1}] which they chose to ignore
(because it vanishes in the dc limit),
and present a detailed discussion of the dependence of the ac
conductance on the various parameters.

As we show, the e-v interaction generates four contributions to
the ac conductance. The first two are a Hartree  and an exchange
terms. In addition,  the Kubo formula for the ac conductance also
contains two vertex corrections. In the Keldysh
formalism,\cite{keldysh, jauho} the vertex corrections correspond
to an expansion of the self energies in the time-dependent
chemical potentials of the left and right reservoirs, $\delta
\mu^{}_{\rm L} (t)$ and  $\delta \mu^{}_{\rm R} (t)$,
respectively. These corrections cause the ac conductance to
diverge  when the frequency $\omega$ of the  applied ac voltage
crosses $\omega^{}_0$; the imaginary part of the Green function of
the vibrational mode is infinitesimally small to  second-order in
the e-v coupling. To avoid this divergence,  a finite lifetime of
the vibrational modes should be included.  Two effects can be
considered which contribute to this lifetime. One  is the
relaxation due to the coupling with bulk phonons in the substrate.
We denote this relaxation rate by $\delta$; the other  involves
the electrical polarization, which results from the coupling of
the vibrational mode with the transport electrons. The latter
effect is dominant for  floating molecules, air-bridged between
the electrodes (in the absence of the substrate).\cite{galperin}
We discuss this case in Sec. \ref{SUMMARY} and in Appendix
\ref{POLA}.

The contributions of  the vertex corrections to the ac conductance
are proportional to  the combination $\Gamma^{}_{\rm L}\delta
\mu^{} _{\rm L} + \Gamma^{}_{\rm R} \delta \mu^{}_{\rm R}$, where
$\Gamma^{}_{\rm L(R)}$ denotes the broadening of the energy
level representing the molecule due to its coupling with the left (right)
lead. Choosing a symmetric configuration,
$\delta \mu^{}_{\rm L}(t)=-\delta \mu^{}_{\rm
R}(t)$ and $\Gamma^{}_{\rm L} = \Gamma^{}_{\rm R}$, causes
these contributions to vanish. Therefore, in
this symmetric case the ac conductance is affected by the e-v interaction only via  the
Hartree and the exchange (Fock) contributions. As we show, the
former contribution is usually dominant, causing a relatively
small increase (larger decrease) of the ac conductance when the
average chemical potential in the leads, $\mu$, is below (above)
the electronic level $\varepsilon^{}_0$ on the molecule. Similar
to the dc conductance, this causes a narrowing of the ac
conductance (plotted versus $\mu$). Since the exchange
contribution is even in $(\mu-\varepsilon^{}_0)$, subtracting the
conductances above and below $\varepsilon^{}_0$ can yield
information on the Hartree contribution.
When the junction is not fully symmetric,
the vertex corrections introduce an additional structure in the
frequency dependence of the ac conductance, especially near
$\omega=\pm\omega^{}_0$. These corrections become maximal in the
limit $\Gamma^{}_{\rm L} = \Gamma$ and $\Gamma^{}_{\rm R} = 0$.

The organization of the paper is as follows. We begin in Sec.
\ref{MODEL} by describing our model and presenting the expression
for the ac conductance of the system. The details of the
derivation are relegated to Appendix \ref{CURDER}. Section
\ref{RES} is devoted to the analysis of the results. In Sec.
\ref{SUMMARY}, we briefly discuss
 proposals for possible
measurements and the effect of the electronic polarization on the
conductance. The random-phase-approximation (RPA) treatment of the
latter effect is outlined in Appendix \ref{POLA}.

\section{The ${\rm ac}$ conductance}

\label{MODEL}

Our model system consists of two electronic reservoirs,
connected together via a single electronic level
$\varepsilon^{}_0$, which represents  the molecule.
The left and right reservoirs
are kept at time-dependent chemical potentials, $\mu^{}_{\rm
L}(t)=\mu+ \delta \mu^{}_{\rm L} (t)$ and $\mu^{}_{\rm R}(t)=\mu +
\delta \mu^{}_{\rm R}(t) $, which oscillate with frequency $\omega$.
When the electron is on the molecule,
it is coupled to a local vibrational mode of frequency $\omega^{}_0$.
This simplified model is described by the Hamiltonian
\begin{align}
\mathcal{H} = \mathcal{H}_{\rm lead}  + \mathcal{H}_{\rm mol}
+ \mathcal{H}_{\rm tun}  .
\end{align}
It consists of the leads' Hamiltonian
\begin{align}
\mathcal{H}_{\rm lead }  =  \sum_{k } (\varepsilon^{}_{k} - \mu^{}_{\rm L} )
c^{\dagger}_{k} c^{}_{ k} + \sum_{p} (\varepsilon^{}_{p} - \mu^{}_{\rm R})
c^{\dagger}_{p} c^{}_{ p},
\end{align}
the Hamiltonian of the molecule
\begin{align}
\mathcal{H}_{\rm mol }  =  \varepsilon^{}_0 c_{0}^{\dagger} c^{}_{0}+ \omega^{}_{0} b^{\dagger}_{}b
+\gamma (b + b^{\dagger}_{}) c^{\dagger}_{0} c^{}_{0},
\end{align}
and
the tunneling Hamiltonian describing the coupling between the molecule and the leads
\begin{align}
\mathcal{H}_{\rm tun} = \sum_{k} (t^{}_{\rm L}
c^{\dagger} _ {k} c^{}_{0}+ {\rm h.c.} )
+ \sum_{p} (t^{}_{\rm R} c^{\dagger}_{p}c^{}_{0} + {\rm h.c.} ).
\end{align}
Here, $c^{\dagger}_{k(p)}$ and $c^{}_{k(p)}$ denote the creation
and annihilation operators of an electron of momentum $k(p)$ in
the left (right) lead, respectively. The creation and annihilation
operators on the level $\varepsilon^{}_{0}$ are denoted by
$c_{0}^{\dagger}$ and $c^{}_{0}$, and  $b^{\dagger}$($b$) creates
(annihilates) a vibrational mode of frequency
$\omega^{}_0$. The coupling of the transport electrons with the
vibrational mode is scaled by $\gamma$. The broadening of
the resonant level on the molecule,
$\Gamma=\Gamma^{}_{\rm L}+\Gamma^{}_{\rm R}$, is given by
$\Gamma^{}_{\rm L (R)}= 2 \pi \nu |t^{}_{\rm L(R)}|^2$, with
 $\nu$ being the density of states of the electrons in the leads.

The current flowing into the molecule from the left  reservoir may
be expressed in terms of the Keldysh Green functions,
\cite{keldysh, jauho}
\begin{align}
I_{\rm L } (t) &= -2e {\rm Re} \int dt^{\prime} \sum_{k}
|t^{}_{\rm L}|^2
                                      \bigl [G^{r}_{00}(t,t^{\prime}) g^{<}_{k}(t^{\prime}, t)
                                       \notag \\
                                      & + G^{<}_{00}(t, t^{\prime}) g^{a}_{k}(t^{\prime}- t)
                                       -  g^{r}_{k}(t- t^{\prime}) G^{<}_{00}(t^{\prime}, t)
                                      \notag \\
                                      \vspace{.2cm}
                                      & - g^{<}_{k}(t, t^{\prime}) G^{a}_{00}(t^{\prime}, t)\bigr ],
                                      \label{eq:cur}
\end{align}
with an analogous expression for the current coming from the right reservoir  (with
L replaced by R and $k$ replaced by $p$).
Here,
\begin{align}
G^{r}_{00}(t, t^{\prime}) & = - i \theta( t - t^{\prime}) \langle \{ c^{}_0(t), c^{\dagger}_0 (t^{\prime}) \}\rangle,
\nonumber\\
G^{<}_{00}(t,t^{\prime})  &=  i  \langle c^{}_0(t^{\prime}) c^{\dagger}_0(t) \rangle,
\end{align}
are the retarded and lesser Green functions on the molecule, where
$\langle \cdots \rangle$ denotes a quantum and statistical
average over the states of the whole system. On the other hand,
the retarded and lesser Green functions on the left (right)
lead,
\begin{align}
g^{r}_{k(p)}(t,t^{\prime})  &= - i  \theta(t-t^{\prime})\langle \{
c^{}_{k(p)}(t), c^{\dagger}_{k(p)}(t^{\prime}) \} \rangle^{}_{0},
\nonumber\\
g^{<}_{k(p)}(t,t^{\prime})  &=  i  \langle
c^{\dagger}_{k(p)}(t^{\prime}) c^{}_{k(p)}(t)  \rangle^{}_{0}.
\end{align}
are given by the average $\langle \cdots
\rangle^{}_{0}$ over the noninteracting leads alone.

\begin{figure}
\begin{center}
\includegraphics[width=6cm]{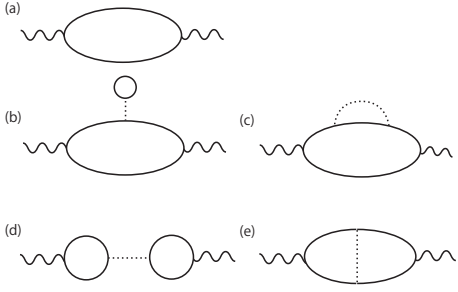}
\caption{ \label{fig:fig1} The diagrams of the ac conductance: (a)
without the e-v interaction,  (b) the Hartree term, (c) the
exchange term,
 (d) the vertex correction for the Hartree term and
 (e) the vertex correction for the exchange term.
The solid line denotes the Green function of the electrons
while the dotted line indicates the  Green function of the vibrational mode.
The wavy line indicates the frequency of the external ac field.}
\end{center}
\end{figure}

Figure \ref{fig:fig1} depicts the diagrams of the  linear-response
ac conductance: (a) in the absence of the e-v interaction,  (b)
including the self energy from the Hartree term and (c) including
the exchange term. Diagrams (d) and (e) are the vertex corrections
of (b) and (c), respectively. The solid line indicates the Green
function of the electron on the molecule whereas the dotted line
denotes the Green function of the vibrational mode. The wavy line
indicates the frequency of the ac field. The diagrams in Fig.
\ref{fig:fig1} are also listed in Ref. \onlinecite{kubala}.
However, the diagram (d) was not calculated in that reference. Our
expressions for the other diagrams agree with those of Ref.
\onlinecite{kubala}.

We expand Eq. \eqref{eq:cur} to first order in $\delta \mu^{}_{\rm
L}(t)$ and $\delta \mu^{}_{\rm R}(t)$ as detailed in Appendix \ref{CURDER}. The
linear-order currents $I^{1}_{\rm L(R)}(\omega)$  are obtained by
substituting the expansions (\ref{eq:epgll}) and (\ref{eq:epgld})
into Eqs. (\ref{eq:gr0}) and (\ref{eq:gl}), and then substituting
the latter expansions into Eq. (\ref{eq:cur}). In our two-lead system, the linear-response
ac conductance can be expressed as a matrix,
\begin{align}
e \begin{bmatrix} I^{1}_{\rm L} (\omega) \\ I^{1}_{\rm R} (\omega) \end{bmatrix}
=\begin{bmatrix}
G^{}_{\rm LL}(\omega) && G^{}_{\rm LR} (\omega) \\ G^{}_{\rm RL}(\omega) && G^{}_{\rm RR}(\omega)
\end{bmatrix}
 \begin{bmatrix}
 \delta \mu^{}_{\rm L} \\ \delta \mu^{}_{\rm R}
 \end{bmatrix}.
\end{align}
We show in Sec. \ref{RES}
that when the molecule is coupled symmetrically to the leads,
$\Gamma^{}_{\rm L} =\Gamma^{}_{\rm R}$,
and the ac voltage is applied symmetrically as well,
$\delta \mu^{}_{\rm L} (\omega )= -\delta \mu^{}_{\rm R}(\omega ) = \delta \mu (\omega )/2$,
the
conductance pertaining to the net current is given by
\begin{align}
G(\omega )& =e\frac{I^{1}_{\rm L}(\omega ) - I^{1}_{\rm R}(\omega )}{2\delta \mu (\omega )}
 = G^{}_{\rm LL}(\omega ) - G^{}_{\rm RR}(\omega ).\label{GSIM}
\end{align}
In this case,  diagrams (d) and (e) (see Fig. \ref{fig:fig1}) vanish since
$\Delta^{}_{\rm 2L}=\Delta^{}_{\rm 2R} =0$ [see Eqs. (\ref{DELTA});   $ \Delta^{}_{\rm 2L(R)} =
\Gamma^{}_{\rm L(R)} (\Gamma^{}_{\rm L} \delta \mu^{}_{\rm L} +
\Gamma^{}_{\rm R} \delta \mu^{}_{\rm R})$]. Consequently, the conductance
given in Eq. (\ref{GSIM}) can be presented as
a sum of three terms,
\begin{align}
G(\omega ) = G^{}_{\rm nint}(\omega ) + G^{}_{\rm H} (\omega )+ G^{}_{\rm ex}(\omega ),\label{GSIM1}
\end{align}
where
\begin{align}
G^{}_{\rm nint}(\omega )& = e \frac{I^{1}_{\rm L-nint}(\omega )- I^{1}_{\rm R-nint}(\omega )}
{2\delta  \mu(\omega )},
\notag \\
G^{}_{\rm H}(\omega )& = e \frac{I^{1}_{\rm L-H}(\omega )- I^{1}_{\rm R-H}(\omega )} {2\delta
\mu(\omega)},
\notag \\
G^{}_{\rm ex}(\omega )& = e \frac{I^{1}_{\rm L-ex}(\omega )- I^{1}_{\rm R-ex}(\omega ) }{2\delta
\mu(\omega)},\label{PARG}
\end{align}
and the partial currents appearing in  Eqs. (\ref{PARG}) are given in Appendix \ref{CURDER},
see Eqs. (\ref{AFT}), (\ref{AFT1}), and (\ref{CONEX}).
The conductance pertaining to the fully-symmetric junction, [see Eqs. (\ref{GSIM}) and (\ref{GSIM1})]
is analyzed in the next section.

In the maximally asymmetric tunneling configuration, where $\Gamma^{}_{\rm L} =\Gamma$ and
$\Gamma^{}_{\rm R} =0$, the conductance of the junction is given by
\begin{align}
G (\omega )=e I^{1}_{\rm L}(\omega )/\delta \mu^{}_{\rm L}(\omega )
    =G^{}_{\rm LL}(\omega ).
\end{align}
In this case there are contributions also from the vertex
corrections, diagrams  (d) and (e) of Fig. \ref{fig:fig1}. We
analyze the conductance of this configuration in Sec. \ref{RES},
expressing it in the form
\begin{align}
G(\omega ) = &G^{}_{\rm nint}(\omega) + G^{}_{\rm H}(\omega )\nonumber\\
& + G^{}_{\rm ex} (\omega )+ G^{}_{\rm verH}(\omega ) + G^{}_{\rm verex}(\omega ).
\label{eq:asy}
\end{align}
Here, each of the partial conductances is given by the respective partial current [see
Eqs.  (\ref{AFT}), (\ref{AFT1}),  (\ref{CONEX}), (\ref{eq:verh}),
and (\ref{eq:verex})], divided by $\delta\mu_{\rm L}(\omega )/e$.

\section{Numerical results for the ${\rm ac}$ conductance}

\label{RES}

As explained in Sec. \ref{INTRO}, one has to allow for a finite
lifetime for the vibrational mode (even at lowest-order in
the coupling $\gamma$), in order to avoid unphysical divergences. In this section
we assume  a constant relaxation rate  (adopting the value $\delta =
0.1 \Gamma$ for the numerical computations); possible contributions of the
 transport electrons to this rate are discussed in the next section.
In the following we measure all energies in units of $\Gamma$, the
broadening of the resonance molecular electronic level brought
about by the coupling to the leads.

 \begin{figure}
 \resizebox{.4\textwidth}{!}
           {\includegraphics{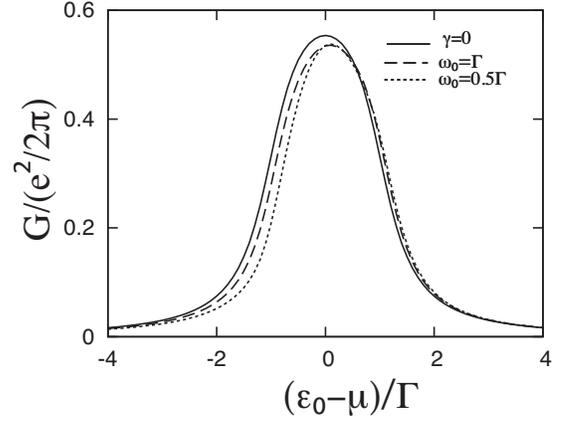}}
 \caption{
 \label{fig:fig2}
 The ac conductance of a  fully-symmetric bridge, as a function of
$ \varepsilon^{}_0 - \mu$ at $\omega = \Gamma$. The coupling strength
of the e-v interaction is $\gamma = 0.3 \Gamma$, and the frequency
of the vibrational mode is     $\omega^{}_0 = \Gamma$ (dashed line) and
$\omega^{}_0 = 0.5 \Gamma$ (dotted line).
 The solid line is the `bare' conductance $G^{}_{\rm nint}$, obtained in the absence of
 the e-v interaction. }
\end{figure}

We begin with the case of a  symmetrically-coupled junction,
$\Gamma^{}_{\rm L} = \Gamma^{}_{\rm R}$.
Figure \ref{fig:fig2} shows the conductance $G$, Eq. (\ref{GSIM}),  as a function of
$\varepsilon^{}_0 - \mu$, for a fixed ac frequency $\omega
=\Gamma$. The solid line indicates $G^{}_{\rm nint}$, the conductance
in the absence of the e-v coupling. It is seen that the
resonance peak does not reach unity as happens in the case of
the dc conductance, \cite{mitra,ora}
due to the suppression by the ac field. When the e-v coupling is
accounted for (the dashed and  dotted lines in Fig. \ref{fig:fig2}) this
peak becomes somewhat narrower, which is
more remarkable for smaller $\omega^{}_0$. This behavior of the
resonance peak is qualitatively similar to the dc case. \cite{ora}

\begin{figure}
 \resizebox{.5\textwidth}{!}
           {\includegraphics{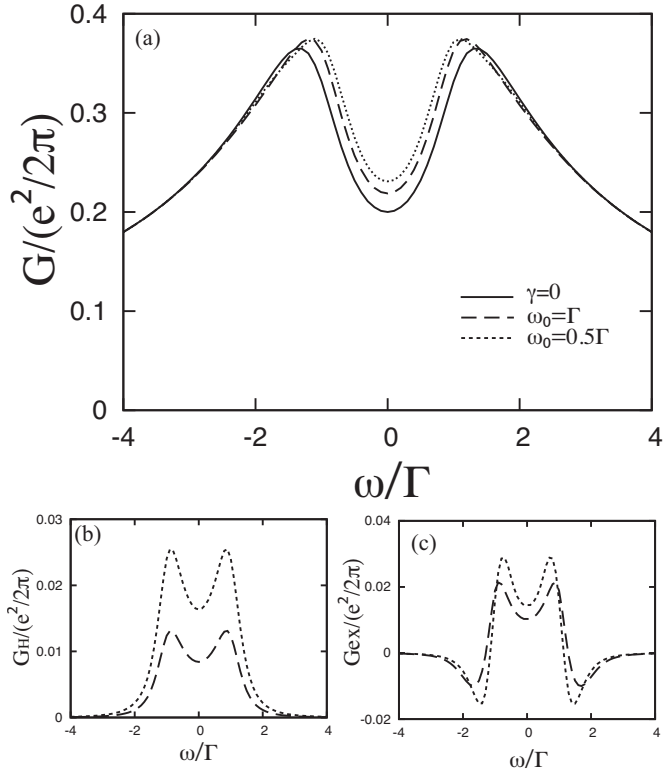}}
 \caption{(a) The ac conductance of  the fully-symmetric bridge, as a function of
 the frequency of the bias voltage $\omega$, at
 $\varepsilon^{}_0- \mu = \Gamma$ and
 $\gamma = 0.3 \Gamma$.
  The frequency of
the vibrational mode is  $\omega^{}_0 = \Gamma$ (dashed line) and
$\omega^{}_0 = 0.5 \Gamma$ (dotted line).  The solid line is the
`bare' conductance $G^{}_{\rm nint}$, in the absence of
 the e-v interaction.  (b) The additional conductance due to
 the Hartree term of the e-v interaction.
 (c) The additional conductance due to the exchange
 term.  \label{fig:fig3}}
\end{figure}

Furthermore, the center of the resonance  peak shifts to a higher energy
as compared to the one in the absence of the e-v coupling. The
narrowing and the shift  may be understood by including   the
Hartree term self-consistently in the electron Green
function  [Eq. \eqref{eq:gr}] employing the Dyson
equation
\begin{align}
G^{r}_{00}(\omega) &= G^{r(0)}_{00}(\omega)
+ G^{r(0)}_{00}(\omega)\Sigma^{r}_{\rm H}(0) G^{r}_{00}(\omega),
\notag \\
&\cong 1/[\omega - \varepsilon^{}_{0} + i\Gamma/2 -
\Sigma^{r}_{\rm H}(0)].
\end{align}
 As can be seen from Eq. \eqref{eq:de2}, $\Sigma^{r}_{\rm H}(0)$ is negative,
and consequently the localized level representing the
molecule is shifted to a lower energy,
implying a lower (higher) conductance for $\varepsilon^{}_0<\mu$
($\varepsilon^{}_0>\mu$). Also the Hartree term renormalizes the
energy scale, which contributes to the narrowing. This
renormalization is due to the polaron binding energy. Notice that the energy
scale is also renormalized by the exchange term,
which oppositely tends to broaden the resonance peak. However, this effect
is smaller than that of the Hartree one (as long as $\delta$
remains small). Since in the linear-response regime there is no real exchange of energy between
the electrons and the vibrational mode, no additional peak
structure appears in the conductance.

\begin{figure}
\begin{center}
\includegraphics[width=8cm]{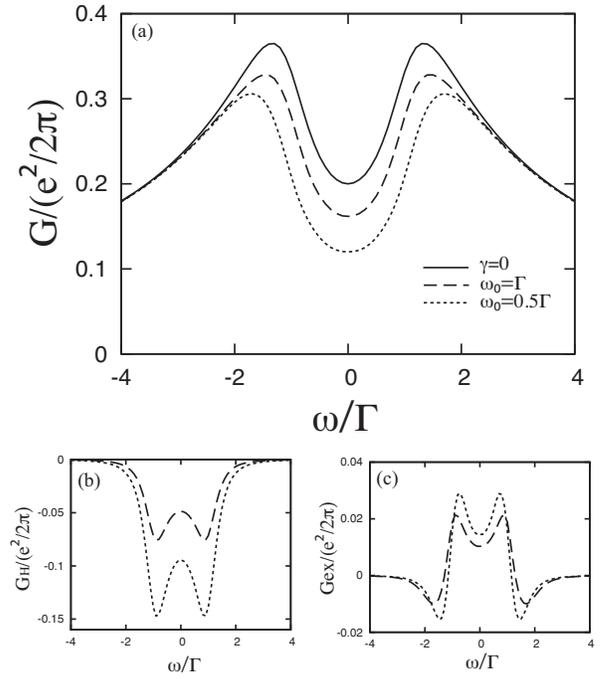}
\caption{
\label{fig:fig4}
The same as Fig. \ref{fig:fig3}, but with $\varepsilon^{}_0 - \mu = - \Gamma$.}
\end{center}
\end{figure}

The full conductance $G$, Eq. (\ref{GSIM}), as a function
of the ac frequency $\omega$ of the external bias when $\varepsilon^{}_0 - \mu =
\Gamma$, is depicted in Fig.  \ref{fig:fig3} (a).
The solid line indicates the conductance in the absence of the e-v coupling, $G^{}_{\rm nint}$.
 Two broad peaks appear around
 $\omega $ of order $ \pm 1.5(\varepsilon^{}_0 - \mu)$.
 The broken lines show $G$ in the presence of the e-v interaction
 with $\omega^{}_0 = \Gamma$ or
 $\omega^{}_0 = 0.5 \Gamma$.
 The e-v interaction increases the conductance
in the region between the original peaks, shifting these peaks to
lower $|\omega|$, while decreasing the conductance  slightly outside this region,
where the e-v effect decays very quickly.
 Figures \ref{fig:fig3} (b) and (c) portray the contributions to the conductance due to the e-v
 interaction coming from the Hartree and the exchange
 terms [diagrams (b) and (c) in Fig. \ref{fig:fig1}],
 $G^{}_{\rm H}$ and $G_{\rm ex}$ respectively, for the same parameters.
 Similar results arise for all positive $\varepsilon^{}_0-\mu$.
 Both $G^{}_{\rm H}$ and $G^{}_{\rm ex}$ show two sharp peaks around
$\omega\sim \pm(\varepsilon^{}_0-\mu)$ (causing the increase in
$G$ and the shift in its peaks), and both decay rather fast
outside this region. In addition, $G^{}_{\rm ex}$ also exhibits
two negative minima, leading to  small `shoulders' in the total
$G$ which are not much visible in Fig. \ref{fig:fig3} (a). The
exchange term virtually shifts the polaron level on the molecule,
yielding an enhancement in the conductance. The amount of increase is more
dominant for lower $\omega^{}_0$.  The situation reverses for
$\varepsilon^{}_0<\mu$, as seen in Fig. \ref{fig:fig4}. Here,
$G^{}_{\rm nint}$ remains as before, but the ac conductance is
suppressed by the e-v interaction. The additional conductance is
dominated by $G^{}_{\rm H}$. The Hartree term of the e-v
interaction renormalizes the energy level in the molecule to lower
values, resulting in the suppression of $G$, while $G_{\rm ex}$
coincides with the case for $\varepsilon^{}_0 - \mu = \Gamma$. The
amount of decrease is larger for lower $\omega^{}_0$. Figures 3
and 4 show that the sign of change in the conductance depends on
the sign of the Hartree contribution,  $G_{\rm H}$.

\begin{figure}
\begin{center}
\includegraphics[width=8cm]{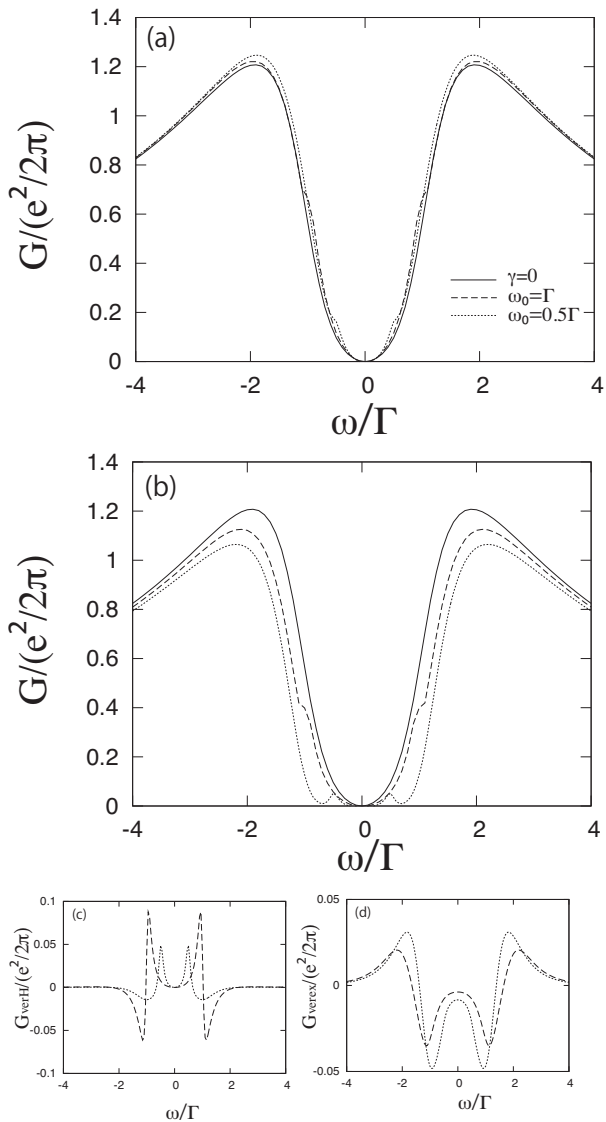}
\caption{ \label{fig:fig5} The ac conductance of an asymmetric
junction, as a function of $\omega$ at (a) $\varepsilon^{} _{0} - \mu=
\Gamma$ and (b) $\varepsilon^{}_{0} - \mu= - \Gamma$ when
$\Gamma^{}_{\rm L} = \Gamma$. The
 energy of the vibrational mode is $\omega^{}_{0} = \Gamma$ (dashed line) or
 $\omega^{}_{0}= 0.5 \Gamma$ (dotted line).
 The solid line is the `bare' conductance $G^{}_{\rm nint}$, in the absence of the e-v interaction.
 The additional conductances resulting  from the vertex corrections of
the Hartree  and   exchange terms are shown in panels (c) and (d).}
\end{center}
\end{figure}

Next, we consider the  conductance Eq. (\ref{eq:asy}) of  an
asymmetrically-coupled bridge for which $\Gamma^{}_{\rm L} =
\Gamma$ and $\Gamma^{}_{\rm R} = 0$. In this case the vertex
corrections are required as explained in Sec. \ref{MODEL}. Figure
\ref{fig:fig5} shows the conductance as a function of the bias
frequency $\omega$  when $\varepsilon^{}_0 - \mu = \Gamma$ (panel
a) and $\varepsilon^{}_0 - \mu = - \Gamma$ (panel b). Similar to
the conductance of a symmetrically-coupled junction
($\Gamma^{}_{\rm L} = \Gamma^{}_{\rm R}$), the ac conductance is
enhanced or suppressed  as compared to the noninteracting case,
depending on whether $\varepsilon^{}_0 > \mu$ or
$\varepsilon^{}_0< \mu$. The contributions to the conductance
arising from the vertex corrections to the Hartree and the
exchange diagrams, $G_{\rm verH}$ and $G_{\rm verex}$  [see Eq.
(\ref{eq:asy})] are shown in panels (c) and (d) respectively. The
vertex corrections coincide for $\varepsilon^{}_0 - \mu = \pm
\Gamma$. Interestingly, the plot (c) exhibits sharp peaks at
$\omega = \pm \omega^{}_0$ due to the singularities of the Green
function of the vibrational mode, which are smeared by its
lifetime.  These also appear as anomalous peaks in the total
conductance. This is due to the charge fluctuation caused by the
ac field, which leads to a fluctuation of the vibrational mode.
This effect becomes blurred when the relaxation rate $\delta$ is
large. Note that the contributions of the vertex correction of the
exchange diagram, $G_{\rm verex}$, cancels some amount of $G_{\rm
ex}$, therefore the enhancement of the conductance is not large
for $\varepsilon^{}_0
> \mu$.

\section{Discussion and Summary}

\label{SUMMARY}

Figures \ref{fig:fig3} and \ref{fig:fig4} show that within our
perturbative expansion the effect of the e-v interaction on the ac
conductance is relatively small and the results do not differ
qualitatively from the expectation without this interaction. One
way to focus on the e-v effect is to compare the conductances at
$\varepsilon^{}_{0} - \mu = \Gamma$ and $\varepsilon^{}_{0} - \mu
= - \Gamma$. As  mentioned in Sec. \ref{RES}, all the contributions to
the conductance except the one coming from the
Hartree term, $G^{}_{\rm H}$, are the same for
$\varepsilon^{}_{0} - \mu = \Gamma$ and $\varepsilon^{}_{0} - \mu
= - \Gamma$. Therefore, one can measure the difference $G^{}_{\rm
H}- G^{\prime}_{\rm H}$ by subtracting the total conductance
$G^{\prime}$ for $\varepsilon^{}_{0} - \mu =- \Gamma$ from $G$ for
$\varepsilon^{}_{0} - \mu =  \Gamma$. Another effect which can be
observable involves the sharp peaks due to the vertex correction
of the Hartree term. This effect is only visible when the
combination $\Gamma^{}_{\rm L} \delta \mu^{}_{\rm
L}+\Gamma^{}_{\rm R} \delta \mu^{}_{\rm R}$ is large and the
relaxation time $\delta$ is small compared to $\Gamma$. Tuning the
tunneling energies $\Gamma^{}_{\rm L(R)}$ can help to extract
these effects from the data.

As stated, the vibrational lifetime for floating molecules
(that are not placed on a substrate) may be
dominated by the effect of the electronic  polarization (i.e., the self energy of
the vibration due to the coupling with the transport electrons), which also
smears the singularities of the Green functions of the vibrational
mode $D^{\alpha}$ ($\alpha$ takes the values $r$, $a$, or $<$) in Eqs.
\eqref{eq:drt} and \eqref{eq:dlt}. In this case, at order
$\gamma^2$,
 $D^{\alpha}$ is  replaced by Eqs. \eqref{eq:drrpa} or \eqref{eq:dlrpa}
in Appendix \ref{POLA}, which include the RPA-type dressing by electrons
shown as the thick dotted line in Fig. \ref{fig:fig6}. Thus, all
the dotted lines in Fig. \ref{fig:fig1} are replaced by the
thicker dotted line. As discussed in Appendix \ref{POLA}, this procedure
generates an additional lifetime for the vibrational modes. For
$\gamma^2/(\omega^{}_0 \Gamma) < 1/4$ we find that the effect of
these RPA-type dressings on the conductance is similar to that of
the effective `ad hoc' relaxation rate $\delta$, as reported
previously.\cite{conf} For larger e-v coupling $\gamma$,  the energy shift
of the electronic level representing the molecule
$\varepsilon^{}_0$ [which corresponds to $\Sigma^{r}_{\rm H}(0)$
in Eq. \eqref{eq:de}] is no longer monotonic as a function of
$\varepsilon^{}_0$. This may imply the breakdown of the
perturbation expansion.

As mentioned in the introduction, and seen in Eqs. (\ref{eq:verh})
and (\ref{eq:verex}), the vertex contributions  [diagrams (d) and
(e) of Fig. \ref{fig:fig1}] appear only for asymmetrically-coupled
bridges. It is diagram (d) of Fig. \ref{fig:fig1} that is most
sensitive to the vibrational mode lifetime, since its contribution
is explicitly proportional to the vibrational Green function (the
dotted line in the center of the diagram). Our results for the
symmetric case, based on diagrams (a)-(c), are indeed not very
sensitive to the vibrational mode lifetime. The RPA treatment of
the electronic  polarization for the case in which the molecule is
coupled {\em asymmetrically} to the leads requires several
additional terms. In addition to diagrams (d) and (e) of Fig.
\ref{fig:fig1}, in the asymmetric case we also need to add
diagrams (f) and (g) of Fig. \ref{fig:fig6}, which represent the
vertex corrections to diagrams (b) and (c) of Fig. \ref{fig:fig1}.
However,  these new contributions involve higher orders of the e-v
coupling $\gamma$, and thus go beyond the scope of the present
paper.

In summary, we have studied the effect of the e-v interaction on
the ac linear conductance of molecular junctions, employing a simple model.  The e-v
interaction enhances or suppresses the conductance, depending on
whether the energy level of the orbital in the molecule is higher
or lower than the chemical potential.
 When the tunnel coupling is asymmetric as defined in Sec. \ref{MODEL} above,
 an additional anomalous structure appears at $\omega = \pm \omega^{}_0$
 due to fluctuations caused by
the ac field.

\begin{acknowledgments}
The authors  thank Y. Utsumi for useful discussions.
This work was partially supported by the German Federal Ministry of
Education and Research (BMBF) within the framework of the
German-Israeli project cooperation (DIP), and by the US-Israel
Binational Science Foundation (BSF).

\end{acknowledgments}

\appendix
\section{Details of the current derivation}

\label{CURDER}

As mentioned, our calculation is carried out to second-order in the
coupling of the transport electrons to the vibrations, $\gamma$.
Thus the Dyson equations for the
Green functions on the molecule
are \cite{mahan}
\begin{widetext}
\begin{align}
&G^{r[a]}_{00}(t, t^{\prime}) = G^{r[a](0)}_{00}(t-t^{\prime})
\notag \\
&+ \int dt^{}_{1} G^{r[a](0)}_{00}(t-t^{}_{1}) \Sigma^{r[a]}_{\rm
H}(t^{}_{1}, t^{}_{1}) G^{r[a](0)}_{00}(t^{}_{1}-t^{\prime}) &+
\int dt^{}_{1} \int dt^{}_{2}
  G^{r[a](0)}_{00}(t-t^{}_{1})
\Sigma^{r[a]}_{\rm ex}(t^{}_{1}, t^{}_{2})
G^{r[a](0)}_{00}(t^{}_{2}-t^{\prime}), \label{eq:gr0}
\end{align}
and
\begin{align}
G^{<}_{00}(t, t^{\prime}) & =
 G^{<(0)}_{00}(t, t^{\prime})
 + \int dt^{}_{1} G^{r(0)}_{00}(t - t^{}_1)
\Sigma^{r}_{\rm H}(t^{}_{1}, t^{}_{1}) G^{<(0)}_{00}(t^{}_{1},
t^{\prime}) +\int dt^{}_{1} G^{<(0)}_{00}(t, t^{}_{1})
\Sigma^{a}_{\rm H}(t^{}_{1}, t^{}_{1})
G^{a(0)}_{00}(t^{}_{1}-t^{\prime})
\notag \\
&+ \int dt^{}_{1} \int dt^{}_{2} G^{r(0)}_{00}(t - t^{}_1)
\Sigma^{r}_{\rm ex}(t^{}_{1}, t^{}_{2}) G^{<(0)}_{00}(t^{}_{2},
t^{\prime}) + \int dt^{}_{1} \int dt^{}_{2} G^{<(0)}_{00}(t,
t^{}_1) \Sigma^{a}_{\rm ex}(t^{}_{1}, t^{}_{2})
G^{a(0)}_{00}(t^{}_{2} - t^{\prime})
\notag \\
& +\int dt^{}_{1} \int dt^{}_{2} G^{r(0)}_{00}(t - t^{}_1)
\Sigma^{<}_{\rm ex}(t^{}_{1}, t^{}_{2})G^{a(0)}_{00}(t^{}_{2} -
t^{\prime}). \label{eq:gl}
\end{align}
\end{widetext}
Here,
$\Sigma^{r}_{\rm H}(t, t)$ represents the self energy
due to  the Hartree term [diagram (b) in Fig. \ref{fig:fig1}],
\begin{align}
\Sigma^{r}_{\rm H}(t, t) = - i \gamma^{2}
\int dt^{\prime}G^{<(0)}_{00}(t^{\prime},{t^{\prime}})D^{r}(t-t^{\prime}),
\label{eq:self-h}
\end{align}
and
$\Sigma^{a}_{\rm H}(t, t) = \Sigma^{r}_{\rm H}(t, t)$. The contribution
of the Fock term [diagram (c) in Fig. \ref{fig:fig1}],
 $\Sigma^{\alpha}_{\rm ex}(t, t^{\prime})$, where
$\alpha$ takes  the values $r$, $a$ or $<$, is
\begin{align}
&\Sigma^{\stackrel{r}{a}}_{\rm ex}(t, t^{\prime}) =  i \gamma^{2} [
G^{<(0)}_{00}(t ,t^{\prime})D^{\stackrel{r}{a}}(t - t^{\prime})
\notag \\
& + G^{\stackrel{r}{a}(0)}_{00}(t-t^{\prime}) D^{<}(t - t^{\prime})
\pm G^{\stackrel{r}{a}(0)}_{00}(t-t^{\prime}) D^{\stackrel{r}{a}}(t - t^{\prime})],
\label{eq:self-exr}
\end{align}
and
\begin{align}
\Sigma^{<}_{\rm ex}(t, t^{\prime}) =  i \gamma^{2}
& G^{<(0)}_{00}(t ,t^{\prime})D^{<}(t - t^{\prime}).
\label{eq:self-exl}
\end{align}
Here,  $G^{\alpha (0)}_{00}(t-t^{\prime})$ are the Green functions
in the absence of the e-v interaction.

The Green functions pertaining to the vibrational mode are given by
\begin{align}
D^{r}(t- t^{\prime})=- i \theta(t  - t^{\prime}) \langle [b^{}(t)
+ b^{\dagger}(t)], [b^{}(t^{\prime}) + b^{\dagger}(t^{\prime})]
\rangle \label{eq:drt}
\end{align}
and
\begin{align}
D^{<}(t- t^{\prime})=-i \langle [b^{}(t^{\prime}) + b^{\dagger}(t^{\prime})],
[b^{}(t) + b^{\dagger}(t)] \rangle.
\label{eq:dlt}
\end{align}
To second-order in $\gamma$, these are required only to order $\gamma^{0}$.
However, as
explained in Sec. \ref{INTRO}, there is a need to assign a finite
lifetime to the vibrations.
Here we  take the relaxation rate (i.e., the inverse lifetime) to be a constant, $\delta$,
arising from a possible coupling to a substrate.
A different scenario is considered in Appendix \ref{POLA}. Thus,
the Fourier transforms of the vibrational Green
functions (at zero temperature) are
\begin{align}
D^{r}(\omega) = \frac{1}{\omega - \omega^{}_0 + i \delta} -
\frac{1}{\omega + \omega^{}_0 + i \delta},\ \  D^{a}(\omega) = [D^{r}(\omega)]^{\ast},
\label{eq:dr}
\end{align}
and
\begin{align}
D^{<}(\omega) = \frac{1}{\omega + \omega^{}_0 + i \delta} -
\frac{1}{\omega + \omega^{}_0 - i \delta}. \label{eq:dl}
\end{align}
Note that all the Green functions except $G^{<(0)}_{00}(t, t^{\prime})$
and $g^{<(0)}_{00}(t, t^{\prime})$ depend only on the time difference $t-t^{\prime}$.

We expand the Green functions pertaining to the electrons to the first order in
the chemical potentials $\delta \mu^{}_{\rm L}(\omega)$
and $\delta \mu^{}_{\rm R}(\omega)$,
\begin{widetext}
\begin{align}
g^{<}_{k(p)}(t, t^{\prime}) & \approx
 i f[\xi_{k(p)}] e^{-i \xi^{} _{k(p)} (t-t^{\prime})}
\exp[ i \int^{t}_{t^{\prime}} dt_{1} \delta \mu^{}_{\rm L(R)}(t)]
= i f[\xi_{k(p)}] e^{-i \xi^{}_{k(p)} (t-t^{\prime})}
 \{ 1 - \int \frac{d \omega}{2 \pi \omega}
[e^{-i \omega t} - e^{-i \omega t^{\prime}}]
\delta \mu^{}_{\rm L(R)} (\omega) \},
\label{eq:epgll}
\end{align}
and
\begin{align}
G^{<(0)}_{00}(t, t^{\prime}) & \approx
\int \frac{d \omega}{2 \pi} e^{- i \omega (t-t^{\prime})}
i f(\omega) \Gamma  G_{00}^{r(0)}(\omega) G_{00}^{a(0)}(\omega)
\notag \\
&+ \int d \omega \int d \omega^{\prime} \frac{i}{(2 \pi)^2 \omega}
[f(\omega + \omega^{\prime}) - f(\omega^{\prime})] e^{- i (\omega
+ \omega^{\prime}) t} e^{i \omega^{\prime} t^{\prime}}
G_{00}^{r(0)}(\omega + \omega^{\prime})
G_{00}^{a(0)}(\omega^{\prime}) (\Gamma^{}_{\rm L}\delta
\mu^{}_{\rm L} + \Gamma^{}_{\rm R} \delta \mu^{}_{\rm R}),
\label{eq:epgld}
\end{align}
\end{widetext}
 where $\xi^{}_{k} = \varepsilon^{}_{k} - \mu$ and
 $f(\omega) = 1/[\exp(\omega/k_{\rm B}T) + 1]$ is
 the Fermi distribution function of the leads.
The Fourier transform of the retarded Green function of the molecule is
\begin{align}
G^{r(0)}_{00}(\omega) = \frac{1}{\omega-\varepsilon^{}_{0} + i\Gamma/2},
\label{eq:gr}
\end{align}
and the advanced Green function is
$G^{a(0)}_{00}(\omega) =[G^{r(0)}_{00}(\omega)]^{\ast}$.

Once the current emerging from the left lead, Eq. (\ref{eq:cur}), is expanded to linear order in
 $\delta
\mu^{}_{\rm L}$ and $\delta \mu^{}_{\rm R}$, it is convenient to divide its Fourier transform
into
five parts,
\begin{align}
I^{1}_{\rm L}(\omega) &= I^{1}_{\rm L-nint} (\omega)
+ I^{1}_{\rm L-H}(\omega) + I^{1}_{\rm L- ex}(\omega)
\notag \\
&+ I^{1}_{\rm L- verH}(\omega) + I^{1}_{\rm L-verex}(\omega).
\label{eq:i1}
\end{align}
The first term on the right-hand side of Eq. (\ref{eq:i1}) is the
current in the absence of the e-v interaction,
\begin{widetext}
\begin{align}
&I^{1}_{\rm L-nint} (\omega) =\frac{e}{2 \pi \omega} \int d
\omega^{\prime} [f(\omega + \omega^{\prime}) - f(\omega^{\prime})]
 {\rm Re} \bigl \{ - i \Delta^{}_{\rm 1L}
[G_{00}^{r(0)}(\omega + \omega^{\prime})
-G_{00}^{a(0)}(\omega^{\prime})]
 + \Delta^{}_{\rm 2L} G_{00}^{r(0)}(\omega +
\omega^{\prime})G_{00}^{a(0)}(\omega^{\prime})  \bigr \}.\label{PRE}
\end{align}
Carrying out the frequency integration, we find
\begin{align}
I^{1}_{\rm L-nint} (\widetilde{\omega}) &=\frac{e\Gamma}{2\pi}
{\rm Re} \bigl ( [( i \widetilde{\omega} -
1)\widetilde{\Delta}^{}_{\rm 1 L} + \widetilde{\Delta}^{}_{\rm 2
L}]\frac{1}{2 \widetilde{\omega} (i + \widetilde{\omega})}
 \{-2 i \arctan[2(\widetilde{\omega}-\widetilde{\varepsilon}^{}_{0})]
 -2 i \arctan[2(\widetilde{\omega}+\widetilde{\varepsilon}^{}_{0})]
 \notag \\
 &+ \log[1 + 4 (\widetilde{\omega} -\widetilde{\varepsilon}^{}_{0})^2]
 + \log[1 + 4 (\widetilde{\omega} + \widetilde{\varepsilon}^{}_{0})^2]
  - 2\log[1 + 4 \widetilde{\varepsilon}^{2}_{0})] \} \bigr ).\label{AFT}
\end{align}
\end{widetext}
We have introduced in Eq. (\ref{PRE}) the notations
\begin{align}
\Delta^{}_{\rm 1L} &= \Gamma^{}_{\rm L} \delta \mu^{}_{\rm L} ,\nonumber\\
\Delta^{}_{\rm 2L} &= \Gamma^{}_{\rm L} (\Gamma^{}_{\rm L} \delta
\mu^{}_{\rm L} + \Gamma^{}_{\rm R} \delta \mu^{}_{\rm R}).\label{DELTA}
\end{align}
In Eq. (\ref{AFT}) and below, we denote by tilde all variables divided by
$\Gamma$ (assigning for them the same notations as before).

The second term on the right-hand side of Eq. (\ref{eq:i1}) is the contribution of the
self energy resulting from the Hartree term $\Sigma^{}_{\rm H}$
[diagram (b) in Fig. \ref{fig:fig1}],
\begin{align}
&I^{1}_{\rm L-H} (\omega) = \frac{e}{2 \pi \omega} \int d
\omega^{\prime}[f(\omega + \omega^{\prime}) - f(\omega^{\prime})]
\notag \\
& \times {\rm Re} \bigl [ (- i \Delta^{}_{\rm 1L}
\{ [G^{r(0)}_{00}(\omega + \omega^{\prime})]^2
-[G^{a(0)}_{00}(\omega^{\prime})]^2 \}
\notag \\
&+ \Delta^{}_{\rm 2L} \{ [G^{r(0)}_{00}(\omega + \omega^{\prime})]^2
G^{a(0)}_{00}(\omega^{\prime})
 \notag \\
 &+ G^{r(0)}_{00}(\omega + \omega^{\prime}) [G^{a(0)}_{00}(\omega^{\prime})]^2 \})
\Sigma^{r}_{\rm H}(0) \bigr ].\label{PRE1}
\end{align}
Here,  $\Sigma^{r}_{\rm H}(0)$ is
\begin{align}
 \Sigma^{r}_{\rm H}(0) =\frac{\gamma^2}{2 \pi} \int^{0}_{- \infty} d \omega
 \Gamma G^{r(0)}_{00}(\omega)
 G^{a(0)}_{00}(\omega) D^{r}(0).
 \label{eq:de}
 \end{align}
Performing the integration in Eq. (\ref{PRE1}) yields
\begin{align}
I^{1}_{\rm L-H} (\widetilde{\omega})& =\frac{e\Gamma}{2 \pi} {\rm
Re} \biggl \{ \frac{32\widetilde{\varepsilon}^{}_{0}[(i
\widetilde{\omega} - 1)\widetilde{\Delta}^{}_{\rm 1L}
+\widetilde{\Delta}^{}_{\rm 2
L}]}{(1+4\widetilde{\varepsilon}_{0}^2)[4\widetilde{\varepsilon}_{0}^2-(i+2\widetilde{\omega})^2]}\biggr
\} \widetilde{\Sigma}^{r}_{\rm H}(0),\label{AFT1}
\end{align}
with
\begin{align}
 \widetilde{\Sigma}^{r}_{\rm H}(0)  =  -\frac{\widetilde{\gamma}^2}{\pi} [\pi -2 \arctan(2 \widetilde{\varepsilon}_0)]
 \frac{\widetilde{\omega}^{}_0}{\widetilde{\omega}_0^2  + \widetilde{\delta}^2}.
 \label{eq:de2}
\end{align}

\begin{widetext}

The contribution of  the exchange terms [diagram
(c) in Fig. \ref{fig:fig1}] to the current is
\begin{align}
&I^{1}_{\rm L-ex}(\omega) =   \frac{e}{2 \pi \omega}
\int d \omega^{\prime} [f(\omega + \omega^{\prime}) - f(\omega^{\prime})]
 {\rm Re}\bigl (-i \Delta^{}_{\rm 1L} [G^{r(0)}_{00}(\omega
+ \omega^{\prime}) \Sigma^{r}_{\rm ex}(\omega + \omega^{\prime})
G^{r(0)}_{00}(\omega + \omega^{\prime})
\notag \\
& -
G^{a(0)}_{00}(\omega^{\prime}) \Sigma^{a}_{\rm ex}(\omega^{\prime})
G^{a(0)}_{00}(\omega^{\prime})]
+ \Delta^{}_{\rm 2L} \{[G^{r(0)}_{00}(\omega + \omega^{\prime})]^2
\Sigma^{r}_{\rm ex}(\omega + \omega^{\prime})
G^{a(0)}_{00}(\omega^{\prime})
+
G^{r(0)}_{00}(\omega + \omega^{\prime})
[G^{a(0)}_{00}(\omega^{\prime})]^2 \Sigma^{a}_{\rm ex}(\omega^{\prime})
\} \bigr ).\label{CONEX}
\end{align}

Finally, the last two terms on the right-hand side of Eq. (\ref{eq:i1}) are
\begin{align}
&I^{1}_{\rm L-verH}(\omega)  =  \frac{e\gamma^2}{(2 \pi)^2 \omega}
\int d \omega^{\prime}\int d \omega^{\prime \prime}
 {\rm Re} \bigl \{ i \Delta^{}_{\rm 2L} [ f(\omega +
\omega^{\prime}) - f(\omega^{\prime})]
 G^{r(0)}_{00}(\omega+\omega^{\prime}) G^{a(0)}_{00}(\omega^{\prime})
D^{r}(\omega)
\notag \\
& \times [ f(\omega + \omega^{\prime \prime}) - f(\omega^{\prime \prime})]
G^{r(0)}_{00}(\omega+\omega^{\prime \prime })
G^{a(0)}_{00}(\omega^{\prime \prime}) \bigr \}
=\frac{e \widetilde{\gamma}^2\Gamma}{(2 \pi)^2 } {\rm Re} \biggr
(\frac{\widetilde{\Delta}^{}_{\rm 2L}} {2\widetilde{\omega}(i +
\widetilde{\omega})^2} \bigl \{ - 2i \arctan[2 (\widetilde{\omega}
- \widetilde{\varepsilon}_0)]
\notag \\
&- 2i \arctan[2 (\widetilde{\omega} +
\widetilde{\varepsilon}^{}_0)]
 + \log[1 + 4(\widetilde{\omega} -
\widetilde{\varepsilon}^{}_{0})^2] + \log[1 + 4(\widetilde{\omega}
+ \widetilde{\varepsilon}^{}_0)^2]
- 2\log [1 + 4  \widetilde{\varepsilon}_0^2] \bigr \}^2
\frac{\widetilde{\omega}^{}_{0}}{(\widetilde{\omega} + i
\widetilde{\delta})^2 - \widetilde{\omega}^{2}_{0}} \biggr ),
\label{eq:verh}
\end{align}
and
\begin{align}
&I^{1}_{\rm L-verex} (\omega)= \frac{e \gamma^2}{(2 \pi)^2 \omega}  \int d \omega^{\prime}
\int d \omega^{\prime \prime}
 {\rm Re} \bigl \{ i \Delta^{}_{\rm 2L} [- f(\omega +
\omega^{\prime}) D^{a}(\omega^{\prime} - \omega^{\prime \prime}) +
f(\omega^{\prime})
 D^{r}(\omega^{\prime} - \omega^{\prime \prime} )
\notag \\
&+ D^{<}(\omega^{\prime} - \omega^{\prime \prime})]
 [f(\omega + \omega^{\prime \prime})
- f( \omega^{\prime \prime})]
 G_{00}^{r(0)}(\omega + \omega^{\prime}) G_{00}^{a(0)}(\omega^{\prime})
G_{00}^{r(0)}( \omega + \omega^{\prime \prime})
G_{00}^{a(0)}(\omega^{\prime \prime}) \bigr \}.
\label{eq:verex}
\end{align}

\end{widetext}

We evaluate the integrals appearing in the expressions for  $I^{1}_{\rm L-ex}$ and $I^{1}_{\rm
L-verex}$ [Eqs.  (\ref{CONEX}) and (\ref{eq:verex})]  numerically as functions of $\widetilde{\omega}$,
$\widetilde{\omega}^{}_{0}$, and $\widetilde{\varepsilon}^{}_{0}$.
The linear expansion of the current emerging from the right lead,
$I^{1}_{R}$, is calculated in the same way.

\section{Green function of the vibrational mode including self energy
corrections due to the transport electrons}

\label{POLA}

Here we solve the Green function of the vibrational mode
taking into account self energy corrections
coming from the coupling with the transport electrons.   We
derive the  electronic polarization induced by those electrons employing
the random-phase-approximation (RPA).

\begin{figure}
\begin{center}
\vspace{.7cm}
\includegraphics[width=8cm]{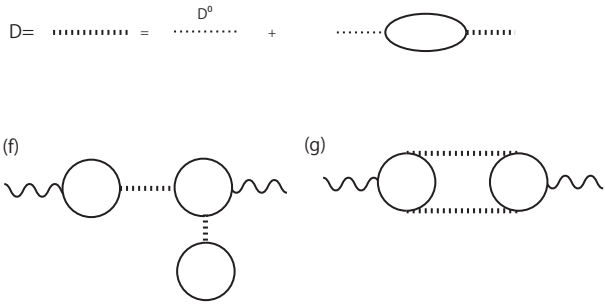}
\caption{ \label{fig:fig6} Top: The thick dotted line denotes the
Green function of the vibrational mode with the RPA term, see
Eqs. \eqref{eq:drrpa} and \eqref{eq:dlrpa}. Bottom: The diagrams
required when the RPA term is included
 besides the diagrams in Fig. \ref{fig:fig1} [with the replacement of
 the dotted line in Fig. \ref{fig:fig1} by the thick dotted line, as shown in the top part].
}
\end{center}
\end{figure}

The relevant diagrams are depicted in Fig. \ref{fig:fig6}, leading to
\begin{align}
&D^{r[a]}(t, t^{\prime}) = D^{r[a](0)}(t-t^{\prime})
\notag \\
& + \int dt^{}_1 \int d t^{}_2 D^{r[a](0)}(t - t^{}_1)
\Pi^{r[a]}(t^{}_1,t^{}_2) D^{r[a]}(t^{}_2, t^{\prime}),
\label{eq:drrpa}
\end{align}
and
\begin{align}
D^{<}(t, t^{\prime}) =\int d t^{}_1 \int dt^{}_2 D^{r}(t, t^{}_1)
\Pi^{<}(t^{}_1,t^{}_2) D^{a}(t^{}_2, t^{\prime}). \label{eq:dlrpa}
\end{align}
Here, $D^{\alpha}$ [see Eqs. (\ref{eq:drt}) and (\ref{eq:dlt})] and $D^{\alpha(0)}$ are the Green
function pertaining to the vibrational mode in the presence and absence
of the coupling to the transport electrons, respectively;
$\Pi^{\alpha}$ is the polarization,
\begin{align}
\Pi^{r} (t, t^{\prime})  = -i \gamma^2
&[G^{<(0)}_{00} (t, t^{\prime}) G^{a(0)}_{00}(t^{\prime}-t)
\notag \\
&+G^{r(0)}_{00} (t-t^{\prime}) G^{<(0)}_{00}(t^{\prime}, t)],
\end{align}
and
\begin{align}
\Pi^{<}(t, t^{\prime}) &= -i \gamma^2 G_{00}^{<(0)}(t, t^{\prime})
 G_{00}^{>(0)}(t^{\prime}, t).
\end{align}
The expressions for the polarization $\Pi^{\alpha}$
contain $G^{<(0)}$, which depends on the chemical potentials in
the reservoirs. Expanding  to linear order in $\delta\mu^{}_{\rm L(R)}$, yields
diagrams (f) and (g) in Fig. \ref{fig:fig6}. This order of the expansion turns
out to be proportional to $\Gamma^{}_{\rm L} \delta \mu^{}_{\rm L}
+ \Gamma^{}_{\rm R}\delta\mu^{}_{\rm R}$, and therefore
diagrams (f) and (g) of Fig. \ref{fig:fig6} do not contribute
to the conductance  of a fully-symmetric junction [Eq. (\ref{GSIM})].
In any event, these diagrams
involve higher orders in $\gamma$, and therefore we discard  them
hereafter.

Confining ourselves for simplicity to the case of a fully-symmetric
junction [see Eq. (\ref{GSIM})],
it is sufficient to keep only
the zeroth order for $G^{<(0)}$, which yields
\begin{align}
D^{r[a]}(\widetilde{\omega}) = \frac{1}{\Gamma} \frac{2
\widetilde{\omega}^{}_{0}}{\widetilde{\omega}^{2} -
\widetilde{\omega}^{2}_{0} - 2 \widetilde{\omega} _{0} \Pi^{r[a]}
(\widetilde{\omega})/\Gamma }
\end{align}
and
\begin{align}
D^{<}(\widetilde{\omega}) = D^{r}(\widetilde{\omega})
\Pi^{<}(\widetilde{\omega}) D^{a}(\widetilde{\omega}),
\end{align}
where
\begin{align}
\Pi^{r}(0)= -\frac{2
\widetilde{\gamma}^2\Gamma}{\pi(1+4\widetilde{\varepsilon}_0^2)}
\end{align}
when $\widetilde{\omega} = 0$. When  $\widetilde{\omega}\ne 0$ we have
\begin{widetext}
\begin{align}
&\Pi^{r}(\widetilde{\omega}) =- i
\frac{\widetilde{\gamma}^2\Gamma}{2 \pi} \bigl [ \frac{1}{2i + 2
\widetilde{\omega}}\{2i \arctan[2(\widetilde{\omega}
-\widetilde{\varepsilon}^{}_0)]
+ 2i \arctan[2(\widetilde{\omega}+\widetilde{\varepsilon}^{}_0)]
\notag \\
&+ 2\ln[1+4\widetilde{\varepsilon}_0^2] -
\ln[1+4(\widetilde{\omega}_{} - \widetilde{\varepsilon}_0)^2] -
\ln[1 +4(\widetilde{\omega} + \widetilde{\varepsilon}^{}_0)^2]\}
\notag
\\
& -\frac{1}{2 \widetilde{\omega} }\bigl (2 i \arctan[2
(\widetilde{\omega} - \widetilde{ \varepsilon}^{}_0)] + 2 i
\arctan[2( \widetilde{\omega} + \varepsilon_0)]
-\ln[1 + 4(\widetilde{\omega}-\widetilde{\varepsilon}^{}_0)^2] - \ln[1 + 4 (\widetilde{\omega} +
 \widetilde{\varepsilon}^{}_0)^2]
+2\ln[1+4\widetilde{\varepsilon}_0^2]\bigr ) \bigr],
 \end{align}
and
 \begin{align}
& \Pi^{<}(\widetilde{\omega})= i \frac{\widetilde{\gamma}^2}{2
\pi} \frac{\Gamma}{\widetilde{\omega} +
\widetilde{\omega}^{3}}\bigl\{2 \widetilde{\omega}
 \arctan[2(\widetilde{\omega} -\widetilde{\varepsilon}^{}_{0})]
 + 2\widetilde{\omega} \arctan[2(\widetilde{\omega}+
 \widetilde{\varepsilon}^{}_0)]
\notag \\
&+ \ln[1+4(\widetilde{\omega} - \widetilde{\varepsilon}^{}_{0})^2]
+ \ln[1+4 (\widetilde{\omega} + \widetilde{\varepsilon}^{}_{0})^2
]
- 2\ln[1+4\widetilde{\varepsilon}^{2}_{0}]
\bigr\}\theta(-\widetilde{\omega}).
 \end{align}
 Clearly, ${\rm Im} \Pi^{r}(\widetilde{\omega})$ and $ \Pi^{<}(\widetilde{\omega})$ yield a
 finite lifetime for the vibrational mode.

\end{widetext}

\end{document}